# The Chinese Pyramids and the Sun

Amelia Carolina Sparavigna
Department of Applied Science and Technology, Politecnico di Torino, Italy

*The Chinese Pyramids are huge ancient burial mounds. In the satellite images we can see some complexes where the main buildings are the pyramidal mounds of an emperor and his empress. Here we discuss a possible sunrise/sunset orientation of these two pyramids on the solstices and equinoxes.*

Chinese pyramids are ancient burial mounds of several early emperors and empresses of China and their relatives. Many pyramids are located near Xi'an, in Shaanxi Province. In this province, in the Lintong district, we find the huge Mausoleum of the First Qin Emperor, Shi Huangdi, accompanied by his large Terracotta Army. The pyramids continued to be built for several centuries, during the following dynasties that ruled China. The shape of the Chinese pyramids is different from those of the ancient Egypt, because they have usually a flat top.

The orientation of the Egyptian pyramids was the subjects of many researches. According to [1-3], the pyramids of the fourth dynasty, at Giza and Dashur, were oriented to face to the cardinal points. About the method used to orient them, in [2] it is told that the precision achieved by the architects was so good that it was certainly based on the rising and setting of stars, not on the measurement of shadows [4]. For the Chinese pyramids, on the contrary, it is not easy to find discussions concerning their orientation. To the author's best knowledge, only two references are available in English; they are an article [5] and the abstract of a conference [6].

The paper of Ref.5 is telling that the pyramids near the cities of Xi'an and Luoyang, together with their suburban fields and roads, are clearly showing a spatial orientation, sometimes along a South-North cardinal direction, sometimes with deviations of several degrees to the East or West. In fact, observing in satellite images these pyramids and the surrounding areas, we can see that the landscape looks like as it had been subjected to a sort of "centuriation" [7]. To determine orientation, Reference 5 is telling that architects and surveyors used a magnetic compass [8]. The needle was directed towards the actual magnetic pole at the times of construction of the tombs. But, magnetic poles shift significantly over time. Therefore the researchers, by matching paleomagnetic observations with some models, identified the date of the pyramid construction in central China.

However, in the abstract [6], astronomical orientations are proposed  The author, Vance Russell Tiede, is telling that two ancient Chinese texts of the Western Han Dynasty, ca. 100 BC, record that the Imperial Astronomer was used to make solar observations to determine the solstices and equinoxes, and for determining the cardinal directions with a circle and gnomon. Probably this ancient astronomer used a method similar to that described by Vitruvius [4]. Moreover, in [6], it is told that, during his investigation of the Chinese pyramids, Tiede determined several astronomical orientation patterns which seems to be not found in the surviving historical record. Let me report a sentence from [6], which is telling that, at the imperial Western Han capital of Ch'ang-an (N 34 latitude), "the diagonals of cardinally oriented square pyramid mounds align to zenith (+34 declination) and nadir (-34 declination) star rise and set points on the skyline". Tiede also observed pyramid orientations to the lunar standstills.

References [5] and [6] are discussing the orientations of a single pyramid. However, in the satellite images, we can see that some burial complexes have two main building, which are the mounds of an emperor and his empress. Using freely available software, sollumis.com [9], we can study the direction of the sun, in particular of sunrise and sunset azimuths on each day of the year, at a specific location visualized on the Google Maps. This software was previously used to easily find the solar orientations of some Syrian stone circles and of the towers of Chankillo, in Peru [10,11]. Let us start searching for a possible sunrise/sunset orientation of the mausoleum of the Emperor Xiaojing of Tang near Goushi, Henan [12], at coordinates 34.63276, 112.81109. The Tang Dynasty ruled during the VII-VIII centuries CE [13]. Figure 1 shows this huge complex, the large pyramid with the flat top is that of the emperor, and near it there is the pyramid of Empress Ai. It is very interesting this group composed by a large pyramid and a satellite pyramid.

Using sollumis.com software, we can immediately see whether a solar orientation exists or not. The result is given in Figure 2. In the upper image, it is shown the direction of the sun on the winter solstice. Observed from the top of the empress' pyramid, the sun is setting over the emperor's pyramid. On the equinoxes, it is possible to see the sun rising and setting between the pyramids, and, on the summer solstice, we could see the sun rising over the empress' pyramid. In particular, from the lower image in the Figure 2, we note that the possibility of finding an observation point on the top of the emperor' mound exists, from which the sun appears to rise over the top of the empress' pyramid.

Are the earlier mounds of emperors and empresses displaying some solar orientations too? We can apply the same software to other complexes having emperor/empress mounds, that we can find in the list of Ref.12. The results are shown in Figures 3 and 4. In the Fig.3, the directions of the sun on winter and summer solstices are shown, as we could see from the pyramids of the Changling and Maoling groups. Note that in the Changling groups (III century BCE), the two mounds have the same size. In that of Maoling of the II century, the empress' mound is smaller. In the Fig.4, we see again the sun on winter and summer solstices over the pyramids of the Duling and Weiling groups, both built during the first century BCE [14]. In the Duling group the two mounds have more or less the same size. In that of Weiling group, the empress' mound is smaller. It seems therefore that there was not a rule governing the relative sizes of the emperor/empress mounds.

In preparing Figures 3 and 4, I chose some possible locations near or over the pyramids, from which it is possible to see the direction of the rays of sun linking the two pyramids. Probably, this fact was considered in the planning of the two mounds, in such a manner that the sun's light could link the consorts in their afterlife. Of course, since the pyramids are quite large, several other locations to observe the sky can be used, changing the overall effect. Therefore, other orientations corresponding to moon standstills or rise and setting of stars can be obtained, besides the proposed solar orientation.

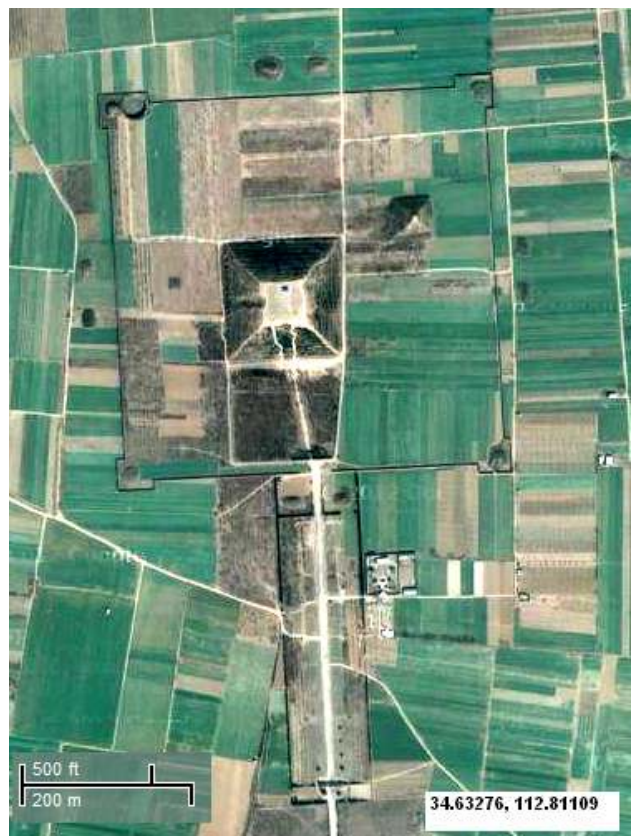

**Fig. 1 The mausoleum of Emperor Xiaojing of Tang near Goushi, Henan. The Tang Dynasty ruled during the VII-VIII centuries CE. The large pyramid with the flat top is that of the emperor, and near it there is the pyramid of Empress Ai.**

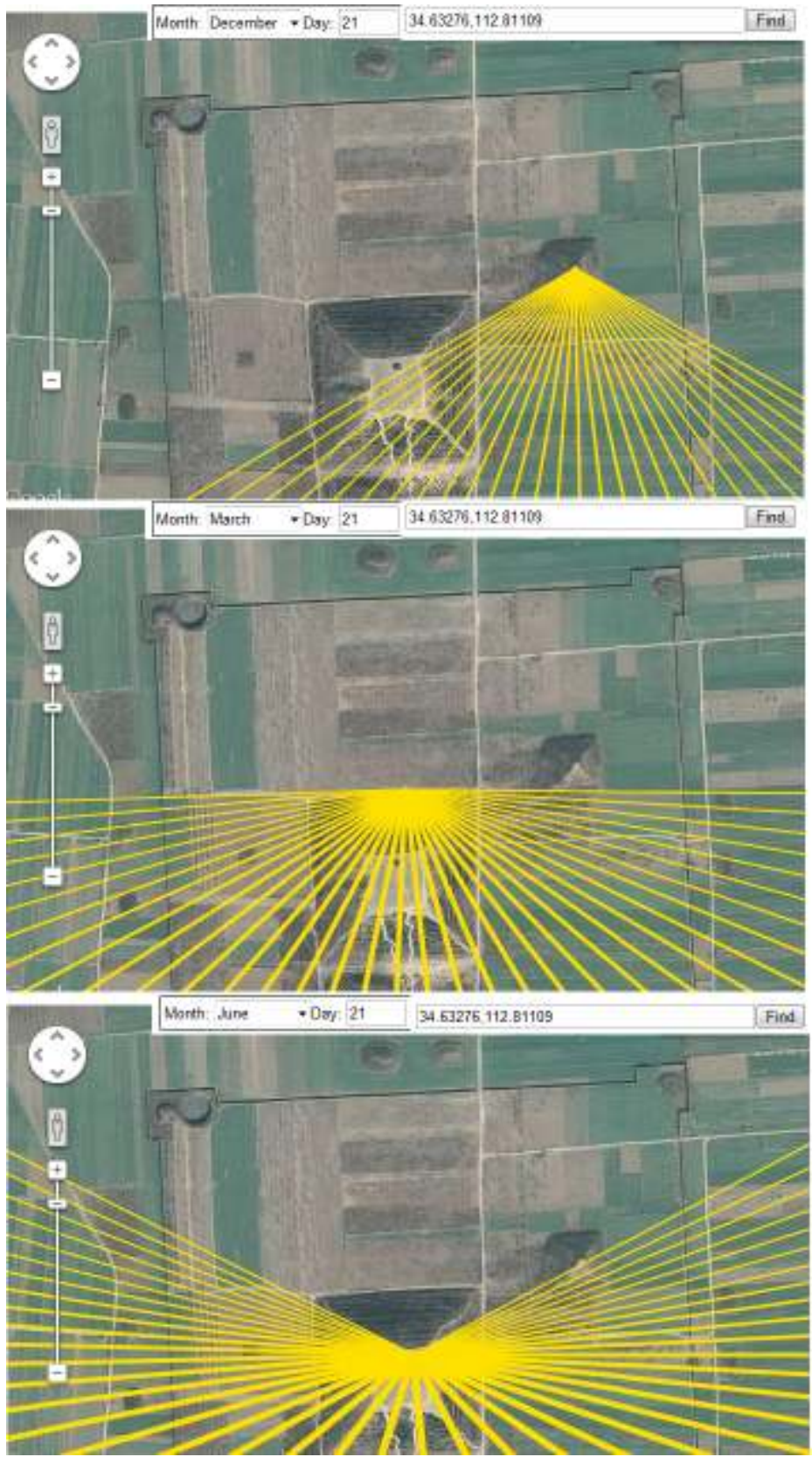

**Fig.2** Using sollumis.com software, we can see a solar orientation of the two mounds. The lines on the drawing show the direction of the sun throughout the day. In the upper panel, it is shown the winter solstice: observed from the top of the empress' pyramid, the sun is setting over the emperor's pyramid. On equinoxes, we could observe the sunrise and sunset between the pyramids, and, on the summer solstice, we could see the sun rising over the top of the empress' pyramid.

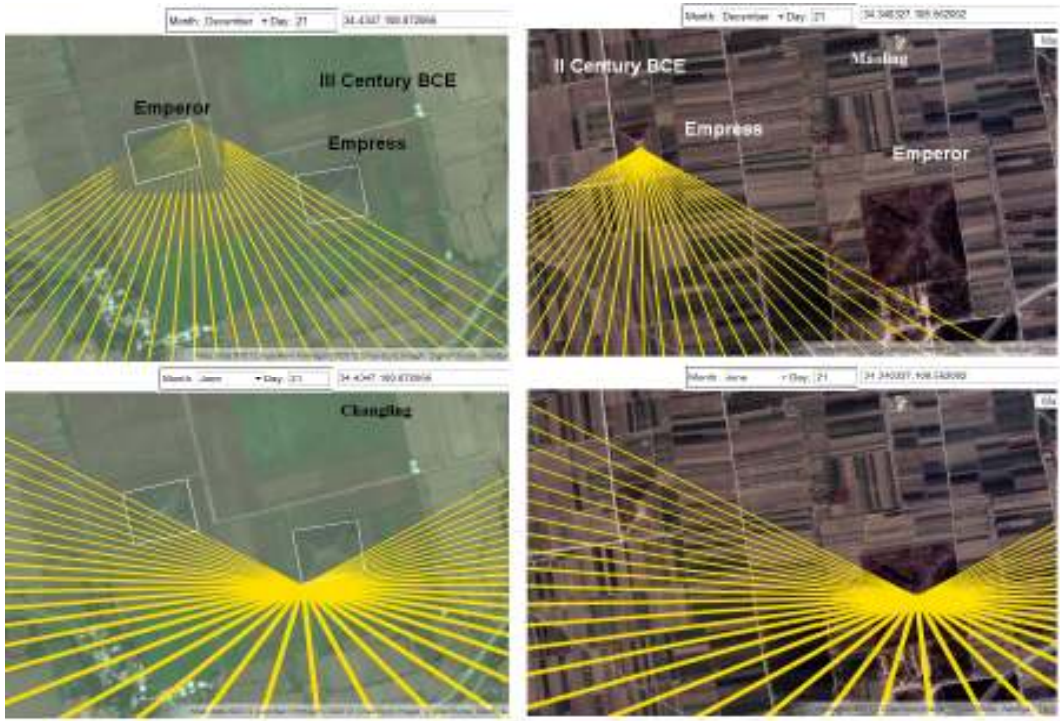

**Fig.3** Sun direction on the winter and summer solstices, as seen from the pyramids of the Changling and Maoling groups. Note that in the Changling groups (III century BCE) the two mounds have the same size. In that of Maoling of the II century, the empress' mound is smaller.

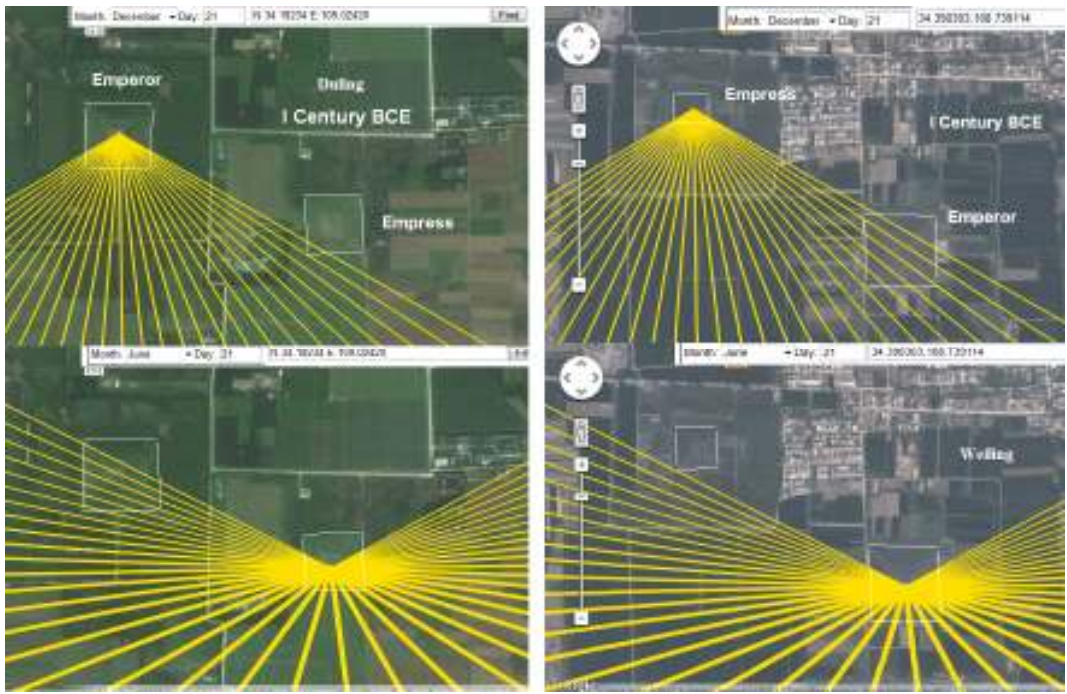

**Fig.4** Sun on the winter and summer solstices, as seen from the pyramids of the Duling and Weiling groups, both of the first century BCE. In the Duling group the two mounds have the more or less same size. In that of Weiling group, the empress' mound is smaller.